\documentclass{PoS}

\title{Grid: A next generation data parallel C++ QCD library}

\ShortTitle{Grid: A next generation C++ library for data parallel QCD}

\author{Peter Boyle, Azusa Yamaguchi, Antonin Portelli\\
       School of Physics, The University of Edinburgh, Edinburgh EH9 3JZ, UK}
\author{Guido Cossu,\\
Theory Center, IPNS,
High Energy Accelerator Research Organization (KEK), Tsukuba, Ibaraki 305-0810, Japan}

\abstract{In this proceedings we discuss the motivation, implementation details, and performance of
a new physics code base called Grid. It is intended to be more performant, more general,
but similar in spirit to QDP++\cite{QDP}.
Our approach is to engineer the basic type system to be consistently fast, rather than
bolt on a few optimised routines, and we are attempt to write all our optimised routines
directly in the Grid framework.
It is hoped this will deliver best known practice performance across the next generation of supercomputers,
which will provide programming challenges to traditional scalar codes.

We illustrate the programming patterns used to implement our goals, and advances in
productivity that have been enabled by using new features in C++11.
}

\FullConference{The 33rd International Symposium on Lattice Field Theory\\
   14 -18 July  2015\\
    Kobe International Conference Center, Kobe, Japan}

\begin{document}

In this proceedings we discuss the motivation, implementation details, and performance of
a new physics code base called Grid. It is intended to be more performant, more general,
but similar in spirit to QDP++\cite{QDP}. Although low level code details are given in
this proceedings, this is intended to convey internal implementation techniques that bring
Grid high performance, and not the high level programming interface itself.


\section{The evolution of computer architectures}

Both engineering constraints and basic physics conspired to ensure that
the modern supercomputer architectures must involve ever growing levels parallelism,
which come in multiple forms.  Common programming paradigms include ad-hoc admixtures
of message passing, threading and single instruction multiple data (SIMD) vector parallelism. 
There is, unfortunately, no good or
universal approach to programming such systems and an ever 
increasing burden is being placed on the programmer to efficiently schedule instructions
to multiple execution units in parallel, see for example Table~\ref{tab:moore}\footnote{only consumer chips with the Skylake architecture are presently available and lack AVX512 support.}.
The Cori, Theta and Aurora systems planned in the (known) US  roadmap all employ many-core processors
with very wide vector lengths. Existing code bases must be reengineered for performance, and we believe this may require
a complete refresh of Lattice QCD codes since, for reasons we explain below, the layout of all data structures must be modified.
\begin{table}[hbt]
\begin{tabular}{ccccccc}
Core & simd &Year & Vector bits & SP flops/clock/core & cores & flops/clock\\
\hline
Pentium III& SSE    &1999 & 128 & 3 & 1 & 3\\
Pentium IV & SSE2   &2001 & 128 & 4 & 1 & 4\\
Core2        & SSE2/3/4 &2006 & 128 & 8 & 2 & 16\\
Nehalem      & SSE2/3/4 &2008 & 128 & 8 & 10 & 80\\
Sandybridge & AVX &2011  & 256 & 16 & 12 & 192\\
Haswell &AVX2     &2013 & 256 & 32 & 18 & 576\\
KNC & IMCI    &2012 & 512 & 32 & 64 & 2048\\
KNL & AVX512  &2016 & 512 & 64 & 72 & 4608\\
Skylake &AVX512 & - & 512 & 64 & - & O(2000)
\end{tabular}
\caption{\label{tab:moore}
The growth in parallel execution resources of one of several semiconductor vendors versus time\cite{Agner,hotchips}. 
Moore's ''law'' for transistor count (while it remains true) guarantees increasing numbers of execution units, 
but it does not guarantee that individual
execution units increase in speed. This is manifest in the growth of flops/clock over time, and the programming
challenge in scheduling O(4000) operations per cycle is clearly inordinately greater than that presented only 15 years ago.
This places a growing burden on the programmer to exploit hardware efficiently.}
\end{table}

In this paper we present a new physics library, called ``Grid'' that employs a high level data parallel approach
while using a number of techniques to target multiple types of parallelism. 
The library currently supports MPI, OpenMP and short vector parallelism. The SIMD instructions sets covered include
SSE, AVX, AVX2, FMA4, IMCI and AVX512. Planned support includes OpenMP 4.0 based offload to accelerators and ARM Neon vector instructions.

As with the QDP++ library\cite{QDP}, and other forms of data parallel programming, the key 
observation is that by specifying high level data parallel constructs in the programming
interface, the details of the subdivision of these operations across underlying forms of parallelism may be obscured and implemented
``under the hood'' thus relieving the programmer of the burden. However, a key difference between our new library and QDP++ is that
both threading and SIMD operations are very efficiently and consistently supported.

It is by now well understood how to approach MPI parallelism in QCD. QCD simulations use regular cartesian (four dimensional) 
meshes, which can be domain decomposed. When the mapping of a simulation grid to the MPI processor grid is uniform, load balancing is perfect
and weak scaling performance is limited only by the halo exchange communication and global reduction operations. Great scalability and
very high application performance can be achieved in systems with high dimensional toroidal or hypercubic topologies\cite{BGQ}.

Hybrid OpenMP and MPI applications are increasingly required to avoid redundant in memory copies when messages are passed internally
in nodes with multiple cores. Further, MPI memory overhead grows with the number of MPI ranks, and this is reduced when hybrid programmes
are used. Existing packages, such as CPS, BFM and QDP++ \cite{CPS,BFM,QDP} have been updated with varying levels of some support for hybrid thread and
message passing since many modern systems such as IBM BlueGene/Q \cite{BGQ,IEEE} and Cray XC30 systems have up to 32 cores per node.

In contrast, updating an existing package to support SIMD parallelisation can be difficult, as we explain in the following sections
where we introduce a suite of techniques and strategies used by Grid to achieve portable performance on QCD codes.

\subsection{SIMD exploitation}

Exploiting locality is a familiar optimisation strategy in the field of computer architecture. The aim is to accelerate the most
common usage cases, while merely correctly executing the less common cases without particular efficiency goals.
Caches, for example, make good use of both spatial locality of data reference (accessing several memory elements in the same cacheline is fast
because a whole cacheline is transferred at once), and of temporal locality of data reference (accessing a memory element repeatedly is accelerated).
SIMD brings a much more restrictive form of locality optimisation. Specifically, the \emph{same} operation must be performed
pairwise on adjacent elements of contiguous (aligned) vectors. The vector length and available operation types vary greatly between architectures.

\section{Grid design patterns}

The first strategy employed by Grid is the \emph{abstraction} of vector operations using modern C++11. Modern compilers
support architecture dependent vector operations using a construct known as ``intrinsic'' functions \cite{intrinsicsguide}. 
These are accessed via  a system dependent header file, such as \verb1<immintrin.h>1. Tying a large code base to such a platform dependent
construct is clearly the mother of all programming evils, and we advance a strategy for doing this in a minimally invasive and easily
retargettable way.

We create C++ vector data type classes, \verb1vRealF1, \verb1vRealD1, \verb1vComplexF1, \verb1 vComplexD1 and \verb1vInteger1. On a scalar architecture
these could simply wrap the corresponding builtin datatypes. Grid localises a modest number of compiler intrinsic sequences
implementing the arithmetic operations for each target for the vector classes in overloaded inline operator functions, enabling
\begin{enumerate}
\item The majority of code to be written using a platform independent interface.
\item The compile time inlining, removal of all functions, and replacement with optimal sequences of assembler operations under reasonable programmer control
\item The vector length varies as dictated by the architecture; all code depending on the vector length access the information from a class member function \verb1 Nsimd()1 and this is used to parameterise optimisations and layout transformations. This contrasts to approaches taken by, for example, the BOOST SIMD library where the vector length is dictated by user code independent of the target architecture\cite{BOOST}.
\item Porting to a new SIMD architecture requires only about 300 lines of code.
\end{enumerate}
We would like to acknowledge useful conversations with a (games) compiler company, CodePlay, located in Edinburgh who advised us that the
games industry has widely adopted these techniques for the encoding of Pauli matrix operations and combined them with the LLVM compiler which
recognises and preserves intrinsic semantics through the optimisation phases. In particular this helps greatly with retaining information in CPU registers.

Semantically, the implementation of the abstracted vector proceeds as follows:
{\small
\begin{verbatim}
#if defined (SSE2)
 typedef __m128 zvec;
#endif
#if defined (AVX1) || defined (AVX2)
 typedef __m256 zvec;
#endif
#if defined (AVX512)
 typedef __m512 zvec;
#endif
 class vComplexD  {
   zvec v;
   // Define arithmetic operators
   friend inline vComplexD operator + (vComplexD a, vComplexD b);
   friend inline vComplexD operator - (vComplexD a, vComplexD b);
   friend inline vComplexD operator * (vComplexD a, vComplexD b);
   friend inline vComplexD operator / (vComplexD a, vComplexD b);
   static int Nsimd(void);
 }
\end{verbatim}}
Here the arithmetic operators are inline functions that, depending on architecture, implement the appropriate sequence of
intrinsic vector functions. We have made reference to the Intel optimisation guides for optimal sequences for complex arithmetic operators.
However, the actual implementation, present in \verb1lib/simd/Grid_vector_types.h1 makes extensive
use of C++11 and type traits: the presentation above is somewhat simplified for the purposes of clarity.

\subsection{Generating optimal small matrix-vector code}

We may now pause to think about the based way to use such vector types for lattice QCD calculations. Since $N_c=3$, one is clearly
facing an impossible challenge to make use of 512 bit vectors in the computation with naturally ordered data composed of $3\times 3$ matrices,
since each contiguous vector is of length 8 single precision complex numbers.

More generally, even if one considers the (irrelevant for QCD, but relevant for finite element codes)   limit where the SIMD vector
length exactly divides the matrix rank, one will still struggle (even with out considering sparse transpose operations) because 
if we consider the naturally optimal approach for matrix $\times$ vector of loading rows of the matrix, and segments of the column vector
into SIMD registers, and accumulating a SIMD vector containing partial contributions to the matrix-vector product, one is left with the 
requirement to horizontally  add the elements of the SIMD summation vector at the end of the row. This horizontal summation sequence
necessarily is a $\log_2$ reduction with serial dependencies requiring the payment of multiple pipeline latency delays.

A key observation in the strategy adopted by Grid is that if we attempt to perform many small matrix-vector products 
at the same time then, providing we interleave the elements of different matrices and the corresponding vectors in consecutive
SIMD lanes, we may perform \verb1 Nsimd() 1 matrix vector products in parallel with perfect SIMD efficiency. 
Using C++ template techniques one can write code that looks the fundamentally the same whether one is performing a single, scalar
matrix-vector product or instead several at a time. Consider the following

{\small
\begin{verbatim}
inline template<int N, class simd>
void matmul( simd *x, simd *y, simd *z)
{
    for(int i=0;i<N;i++){
        for(int j=0;j<N;j++){
            x[i] = x[i]+y[i*N+j]*z[j];
        }
    }
}
\end{verbatim}
}
Here, the underlying type \emph{simd} is a template parameter. One could pass in the template parameter \verb1 simd = float 1
and obtain a standard scalar matrix-vector multiply. On the other hand, passing in a Grid vector type as the template
parameter has the effect of \emph{widening} the basic datum to 128, 256, or 512 bits depending on architecture and the
\emph{same} routine will equally and automatically perform four, eight or sixteen matrix vector products in parallel.
There is no need to horizontally add across SIMD lanes, and no impact of pipeline latency beyond any seen by scalar code.
The vector code is essentially identical to scalar code, except that \emph{sizeof(double)} appears to have grown
to 128/256/512 bits. As a result the routine is automatically 100\% SIMD efficient.

\begin{figure}[hbt]
\includegraphics[width=0.45\textwidth]{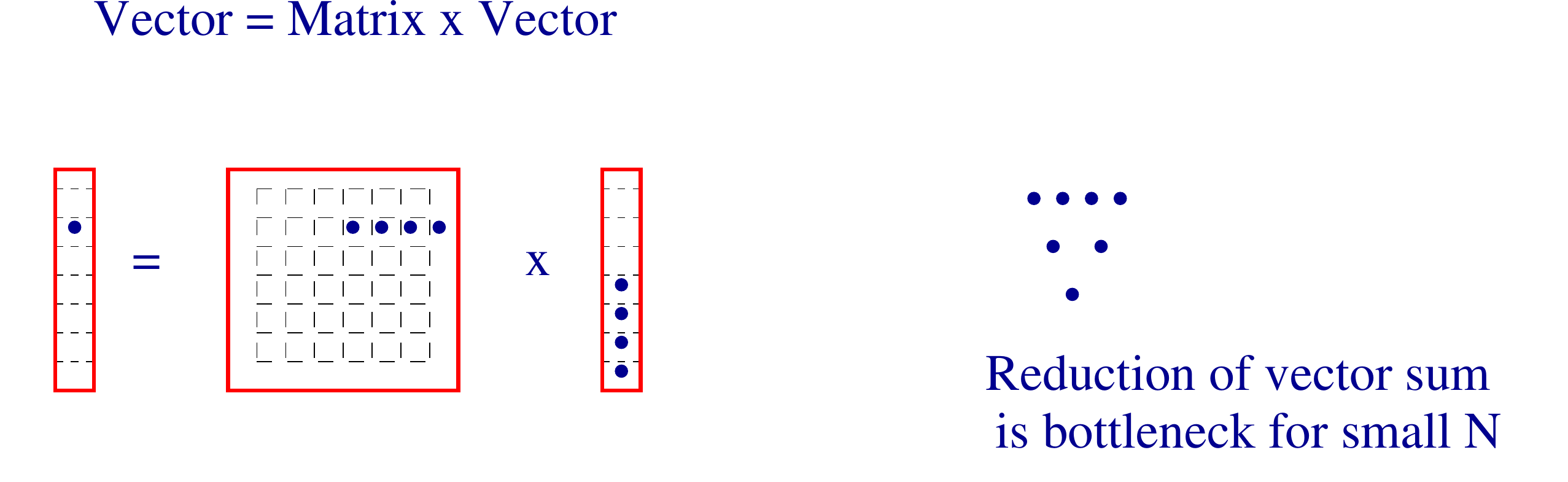}
\includegraphics[width=0.45\textwidth]{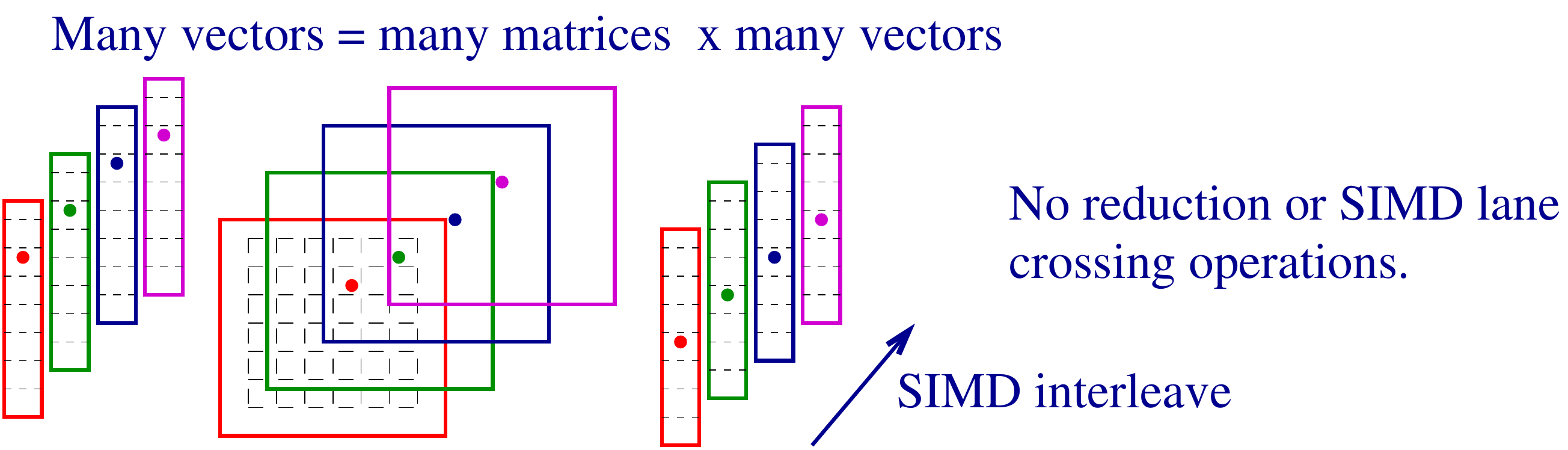}
\caption{\label{fig:mvec} Two different approaches to SIMD accelerating matrix-vector products. The left approach
suffers from long latency operations when horizonally adding a summation register (in addition to the
fact that increasing vector lengths do not typically neatly divide the index ranks of QCD fields), while the right hand approach
to performing many of these operations does not. 
}
\end{figure}

Further, the integer template parameter $N$ is known at compile time. Loops are automatically unrolled by modern versions
of GCC, Clang/LLVM and Intel's compilers. Evidence will presented later demonstrating that high fractions of peak performance,
of order 65\% on cache resident data, are achievable on modern Intel processors.

\subsection{Data parallelism and cartesian layout transformation}

Exposing arbitrary wide levels of appropriately aligned SIMD operations is a challenge. However it has a historical (but
non-obvious solution). As long ago as the 1980's data parallel languages such as Connection Machine
Fortran and High Performance Fortran exposed high level array layout transformations in the programming interface. 
The CM-2 is of particular interest\cite{CM2}; specifically geometrical domain decomposition was performed over more \emph{virtual} nodes than there
were physical nodes. The mapping of array indices to virtual nodes was controlled at compile
time through \emph{layout} directives (a \emph{cmfortran} language extension). We aim to apply similar techniques in modern C++.

\begin{table}[hbt]
\begin{tabular}{cccccc}
ISA & vRealF & vRealD & vComplexF & vComplexD & default layout\\
\hline
SSE    & 4 & 2 & 2 & 1 & 1.1.1.2\\
AVX    & 8 & 4 & 4 & 2 & 1.1.2.2\\
AVX512 & 16& 8 & 8 & 4 & 1.2.2.2
\end{tabular}
\caption{\label{tab:layout}
Vector lengths for different data types used by Grid under different compile targets, and the corresponding
default virtual node over decomposition (in x,y,z,t order) for single precision complex fields.}
\end{table}

\begin{figure}[hbt]
\includegraphics[width=0.5\textwidth]{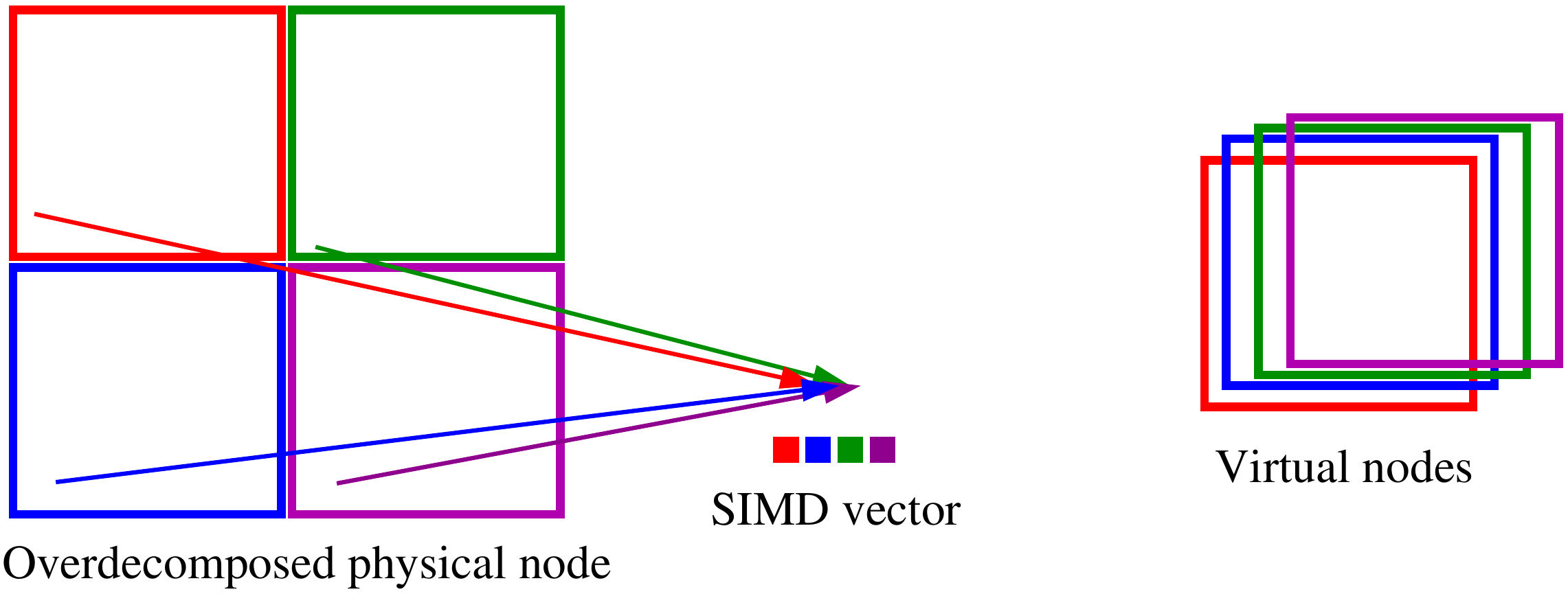}
\caption{\label{fig:overdecompose} 
A key idea is to overdecompose and then interleave elements from different virtual nodes in adjacent SIMD lanes. After transformation
it is simple to generate code such that one node performs the work on $N=4$ virtual nodes, each with a correspondingly reduced local volume.
}
\end{figure}

Some aspects of the above idea may be reused. By over-decomposing the geometrical subdivision of a Cartesian array problem
into smaller problems than suggested by the MPI task count (in this case by a factor of four) a layout transformation
can be generated that allows us to obtain 100\% SIMD efficiency \cite{BFM,BGQ}, Figure~\ref{fig:overdecompose}.
Grid overdecomposes any given cartesian grid over virtual nodes, appropriate
to filling the SIMD vector lengths, as shown in Table~\ref{tab:layout}.
As with QDP++, distributed arrays are held in Lattice container classes whose internal layout is hidden from higher level code.
This abstraction of layout is essential as it enables us to \emph{transform} the layout in an architecture dependent and optimal way.
\begin{itemize} 
\item Multiple \emph{Grid} objects may be active, each loosely corresponding  to the single \emph{Layout} namespace in QDP++. When a
field is constructed, a Grid object must be passed. 
\item Grid objects can be of any dimension, and contain the information of decomposition across MPI tasks, OpenMP thread counts, and SIMD lanes.
\item Four dimensional and five dimensional Grids, multiple levels of coarse and fine Grids, and single checkerboard Grids may be active concurrently
      in the programme.
\item \emph{conformable} array operations between fields on the same grid proceed data parallel with good SIMD efficiency. Array slicing is not
      supported but there are a limited set of grid transferral operations appropriate to a modern, multi-grid aware QCD code.
\end{itemize}

\subsection{Cshifting implementation}

Circular shift of Grid lattice containers by an arbitrarily large distance in any dimension is supported.
For example, the following code performs a data parallel multiplication a distributed fields B and C together and takes the y-derivative of C,
with work being decomposed across MPI ranks, OpenMP threads and SIMD lanes automatically.
{\small
\begin{verbatim}
LatticeColourMatrix A(Grid), B(Grid), C(Grid), dC_dy(Grid);
int Ydim = 1;
       A = B*C;
   dC_dy = 0.5*( Cshift(C,Ydim,+1) - Cshift(C,Ydim,-1) );
\end{verbatim}
}

\begin{figure}[hbt]
\begin{center}
\includegraphics[width=0.8\textwidth]{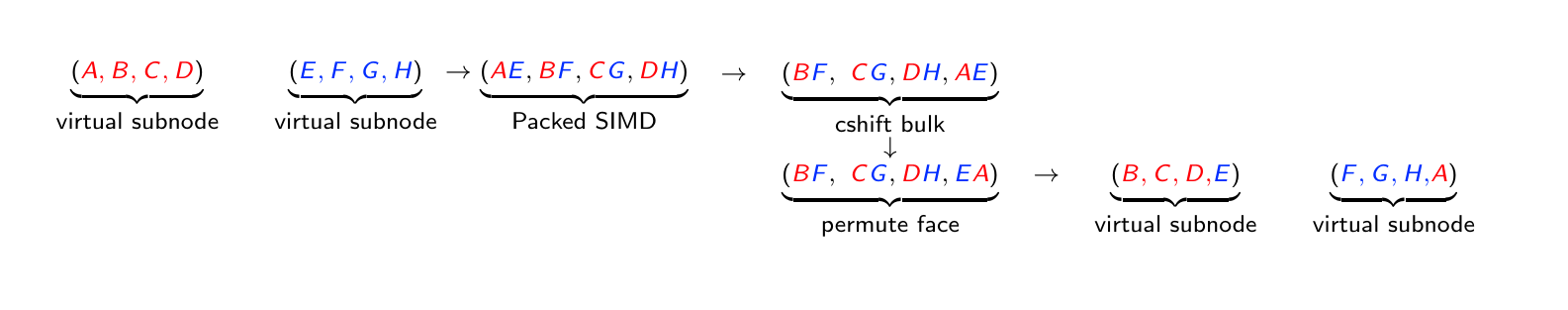}
\end{center}
\caption{\label{fig:cshift} Permutation, and all SIMD lane crossing overhead, is suppressed by the surface to volume ratio in our
interleaved virtual node scheme; we illustrate here a one-dimensional example of a cshift operation. Only the interior
three steps are performed in grid since this is the native, internal data layout -- the exterior diagrammatic representation as
unpacked into virtual nodes is illustrative only.}
\end{figure}

While, the SIMD efficiency of site local operations is clear in our scheme, for site non-local operations such as a finite difference
operator this is also true but less clear. The symmetry of a cartesian mesh means that the \emph{neighbours} of any element
of one virtual node also align with the neighbours of corresponding elements of other virtual. 
The code for processing $N$-virtual nodes therefore looks identical to that for processing a single
scalar node but the datum is simply $N$-fold larger. 
For stencil connections that cross boundaries
between sub-cells a permutation, is additionally required since for example a blue element
of the result vector may need to reference a red neighbour and vice versa, see Figure~\ref{fig:cshift}.
These non-arithmetic steps are both
efficient SIMD operations, \emph{and are suppressed by the surface to volume ratio}.

\subsection{Stencil support}

A Cshift closes an expression template, to avoid complex implicit book keeping in expressions such as a \emph{cshift} of a \emph{cshift}.
It is wise to make it easy to implement high performance routines for PDE matrix stencils: we provide a generic stencil assistance. Our quoted
performance figures are from codes are implemented using this strategy. We give below the example of a a coarse grid operator used
in an adaptive multigrid QCD inverter.

\begin{minipage}{0.5\textwidth}
\includegraphics[width=\textwidth]{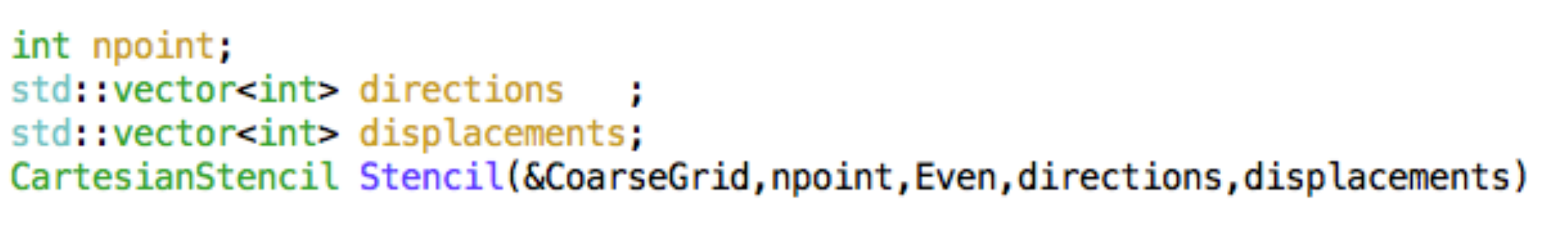}\\
\includegraphics[width=0.9\textwidth]{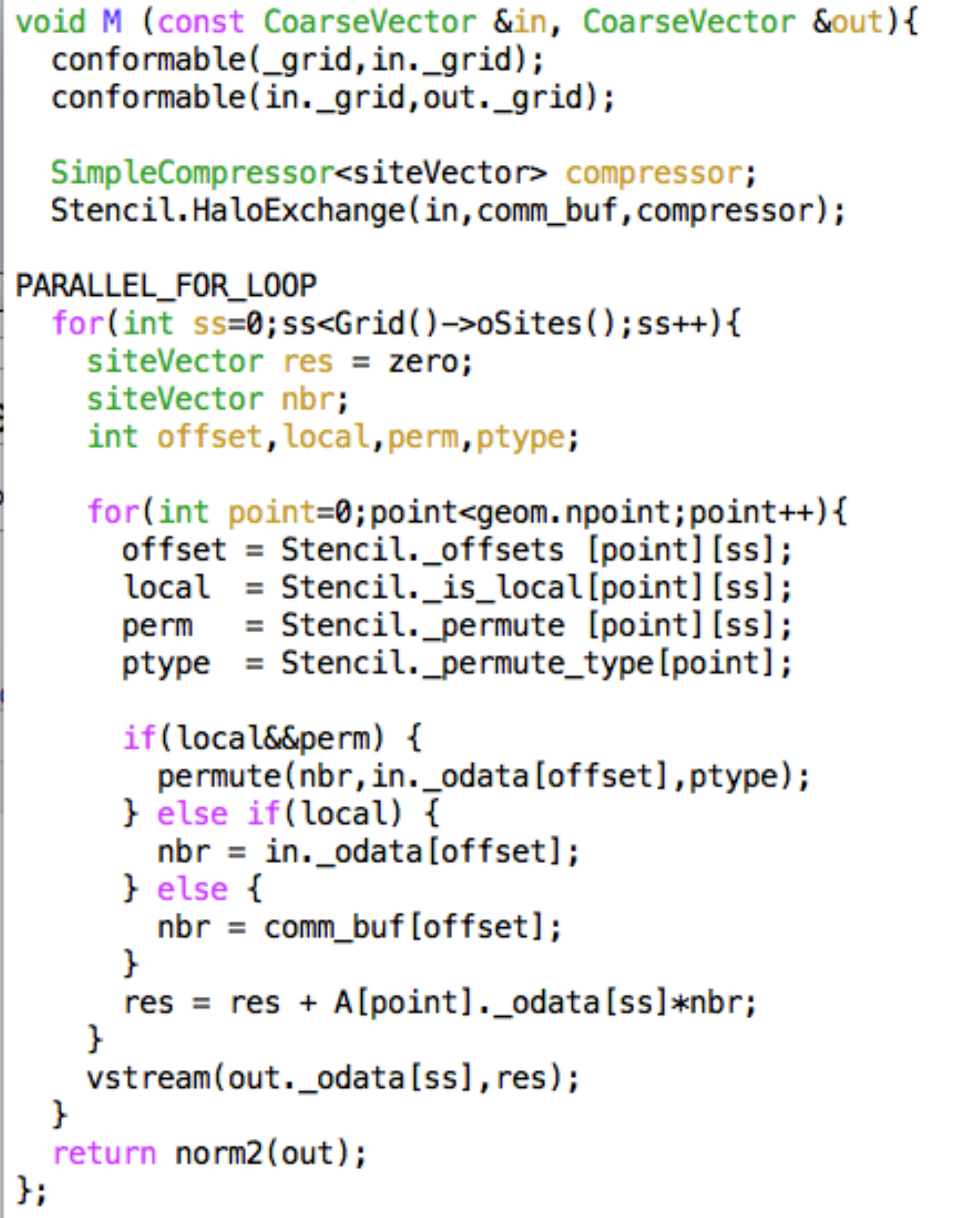}
\end{minipage}

The geometrical structure in grid dimensions is specified \emph{with no reference} to the type of field being operated on. It could
equally be a complex scalar field, or a rich spinor from the perspective of the Stencil class which is concerned only with the finite
difference geometry. They details of the field types are passed into the HaloExchange method as a template, and the data motion
is performed by Halo exchange for whatever type is exchanged. A \emph{Compressor} functor is used, which may change precision (or spin project)
communicated fields, and once the halo exchange is complete a node has only local computation to perform. 
The local computation is performed by the PDE kernel writer, with indexing and permutation performed as advised by the Stencil geometry object.
Streaming stores are used in the final write to memory. This code captures the rather longer but similar idioms used in our Wilson Kernel implementation.

\subsection{Tensor type structures \& automating layout transformation}

We make use of C++11 features to reduce the volume of code and increase the flexibility compared to QDP++.
We implement internal template classes representing Scalar, Vector or Matrix of \emph{anything}.
{\small\begin{verbatim}
template<class vtype> class iScalar
{
    vtype _internal;
};
template<class vtype,int N> class iVector
{
    vtype _internal[N];
};
template<class vtype,int N> class iMatrix
{
    vtype _internal[N][N];
};
\end{verbatim}
}
The type templating is useful in several ways:
\begin{enumerate}
\item Passing scalar objects such as RealF, RealD, complexF or ComplexD produces the normal scalar objects one would naturally define.
\item We can pass in the types vRealF, vRealD, vComplexF, or vComplexD. All our mathematical data structures and functions that operate
      on them reorganise to adopt a the number of virtual nodes per physical node that is consistent with the architectures vector width.
      Further the code for data parallel arrays is \emph{the same code} as our scalar implementation. Modern C++ should indeed be 
      be viewed as primarily a system for strictly minimising the number of independent places the same functionality need be implemented.
\item The Scalar, Vector and Matrix types may be passed as template parameters to each other, building up arbitrary tensors.
For example, \\
\verb1 Vector<Vector<Vector<RealF,Ncolour>, Nspin >, Nflavour > 1\\
could represent a flavoured Fermion field in a G-parity simulation.
\end{enumerate}
In particular the basic library does not fix the number of indices in the tensor product structure of our types.
The key element is to teach the system how to apply the multiplication table for any tensor product combination of
scalar, vectors and matrices. 

Here, for example, the QDP++ approach based on C++ '98 made use of the now
discontinued PETE expression template engine to \emph{enumerate all possible cases} in O(60k)lines
of machine generated code. 
With the introduction of \emph{auto} variables, and the recognition of the \emph{decltype} of an expression 
we can do rather better, and recursively infer the return type for arithmetic operators on $n$-deep nested tensor
objects from
\begin{enumerate}
\item the return type of the same operation on the $n-1$ index nested object
\item the multiplication rules for scalar/vector/matrix on the $n$-th index
\end{enumerate}
\begin{table}[hbt]
\begin{tabular}{cccc|cccc|cccc}
L & op & R &$\to$ ret &
L & op & R &$\to$ ret &
L & op & R &$\to$ ret \\
\hline
S & $\times$ & S & $\to$S &
S & $\times$ & V &$\to$ V &
S & $\times$ & M &$\to$ M \\
V & $\times$ & S &$\to$ V &
V & $\times$ & V &$\to$ S &
V & $\times$ & M &$\to$ V \\
M & $\times$ & S &$\to$ M &
M & $\times$ & V &$\to$ V &
M & $\times$ & M &$\to$ M 
\end{tabular}
\caption{\label{tab:multtab} Elemental rules are implemented for our scalar matrix and vector classes. We may use these
and modern C++11 syntax to infer the rules for composite types in arbitrary tensor combinations. For clarity,
if there were five index types in composition the C++ 98 approach would require the $9^5$  cases to be enumerated in
inline functions in header files and this is a combinatorially large saving in code volume.}
\end{table}
We must then only build up the elemental multiplication table, Table~\ref{tab:multtab} and similar tables
for addition, subtraction etc. An example of the syntax is given below, and it is worth noting the
implemetation using \emph{mult} and \emph{mac} routines both enables the use of fused multiply add sequences,
passes pointers protected by the \emph{restrict} key word for performance, and passes a return pointer to eliminate object copyback.
{\small
\begin{verbatim}
template<class l,class r,int N> inline
auto operator * (const iMatrix<l,N>& lhs,const iVector<r,N>& rhs)
       -> iVector<decltype(lhs._internal[0][0]*rhs._internal[0]),N>
{
 typedef decltype(lhs._internal[0][0]*rhs._internal[0]) ret_t;
 iVector<ret_t,N> ret;
 for(int c1=0;c1<N;c1++){
   mult(&ret._internal[c1],&lhs._internal[c1][0],&rhs._internal[0]);
   for(int c2=1;c2<N;c2++){
     mac(&ret._internal[c1],&lhs._internal[c1][c2],&rhs._internal[c2]);
   }
 }
 return ret;
}
\end{verbatim}
}

\subsection{Expression template engine}

The Grid library consistently instantiates the tensor classes with SIMD types when dealing with cartesian coordinate
dependent fields, and scalar types for site scalars. The data parallel fields are wrapped in opaque Lattice container classes
so that the programmer will mainly deal with high-level data parallel constructs. 

It is common to implement expression template engines, to enable the cache friendly fusion of operations in a single
site (vector) loop, to avoid therepeated application of long vector binary operations in compound expressions
\cite{Blitz}. The technique was introduce to the Lattice QCD domain by QDP++\cite{QDP}.
Using new elements of C++11 syntax we have been able to reduce the implementation of the Grid lattice expression template
engine to under 350 lines of code.
For example, binary expressions of lattice container objects can infer their return type from the corresponding expression
assembled from the objects \emph{contained} within the container class and the many cases do not need to be enumerated,
giving several orders of magnitude reduction in code.

The closure of all lattice object expresion templates includes an OpenMP threaded pragma so that threading is automatic.
It is intended that in future OpenMP 4.0 offload pragmas will be made an option and a model where all parallel array
statements are processed on the accelerator, while scalar operations are processed on the host will be followed. This
is notably similar to the connection machine approach.

\subsection{Object serialisation and liberation from XML} 

We have implemented a variadic macro based system that will automatically generate calling code
for a set of virtual reader and write classes for parameter classes, as part of the class definition. 

This removes duplicative the chore of generating XML (and other format) I/O code for many parameter classes, 
and provides access to multiple file representations due to the virtualisation of the scheme. We use the PugiXML
library which is both small enough (single file) and licensed to be included along with the source code and this
avoids dependency on an external library for XML parsing, with a minimal coupling of the code to any particular
file format.
As an example, 
{\small
\begin{verbatim}
class myclass {
public:
  GRID_DECL_CLASS_MEMBERS(myclass,
                          int, x,
                          double, y,
                          bool , b,
                          std::string, name,
                          std::vector<double>, array);
};
\end{verbatim}
}
both declares a serialisable class, and generates flexible serialisation code. The macro is implemented
in \emph{lib/serialisation}. This is not disimilar to an approach introduced in the \emph{BOOST} Hana\cite{BOOST}
library\footnote{we thank Robert Mawhinney for pointing the Hana implementation out to us, which we
read, and ultimately adopted our own independent implementation.}, but our scheme
is more general and avoids dependency on a library.

\section{Performance} 

\begin{figure}[hbt]
\includegraphics[width=0.8\textwidth]{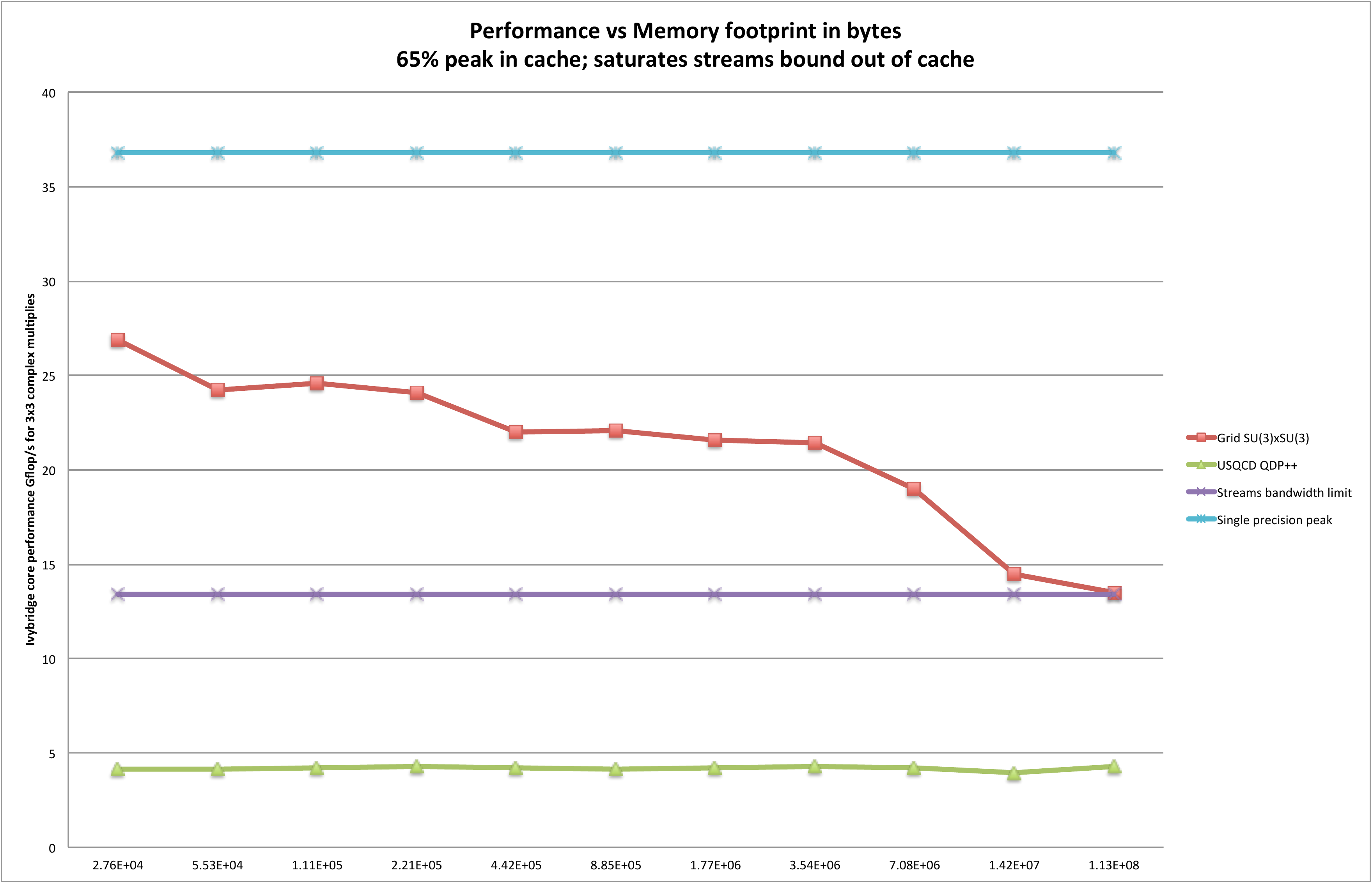}
\caption{\label{fig:su3} We compare the performance of Grid (red) on SU(3)$\times$SU(3) matrix multiplication to peak (blue), 
the limit imposed by memory bandwidth (purple), and to that of the QDP++ code system (green).
}
\end{figure}

We have made claims that our code system removes a large amount of C++ sophistication to inline SIMD intrinsic functions 
and deliver better than fortran performance. These claims must now be substantiated! 
First we analyse a simple kernel forming the product of fields of complex SU(3) matrices.
The benchmarked code is 
{\small
\begin{verbatim}
      LatticeColourMatrix z(&Grid);
      LatticeColourMatrix x(&Grid);
      LatticeColourMatrix y(&Grid);
      z=x*y;
\end{verbatim}}
The Grid is varied to map the single precision performance as a function of memory footprint in Figure~\ref{fig:su3} on
a single core of an Intel Core i7-3615QM 2.3GHz CPU using the Clang v3.5 compiler. This represents 65\% of peak performance 
(2$\times$2.3$\times$256/32 = 36.8GF peak) when 
data is resident in L2 cache. When data is not present in L2 cache the STREAM bandwidth is slightly exceeded by Grid. 
We compare
to the QDP++ implementation using g++ v4.9, and no SSE kernels enabled so that the performance of template type systems are being compared.
QDP++ is around 5.5 times slower on the same chip for the same operation. QDP++ did not compile with the Clang compiler and did not
particularly benefit from the cache since it was not saturating memory bandwidth. 

\begin{figure}[hbt]
\includegraphics[width=0.5\textwidth]{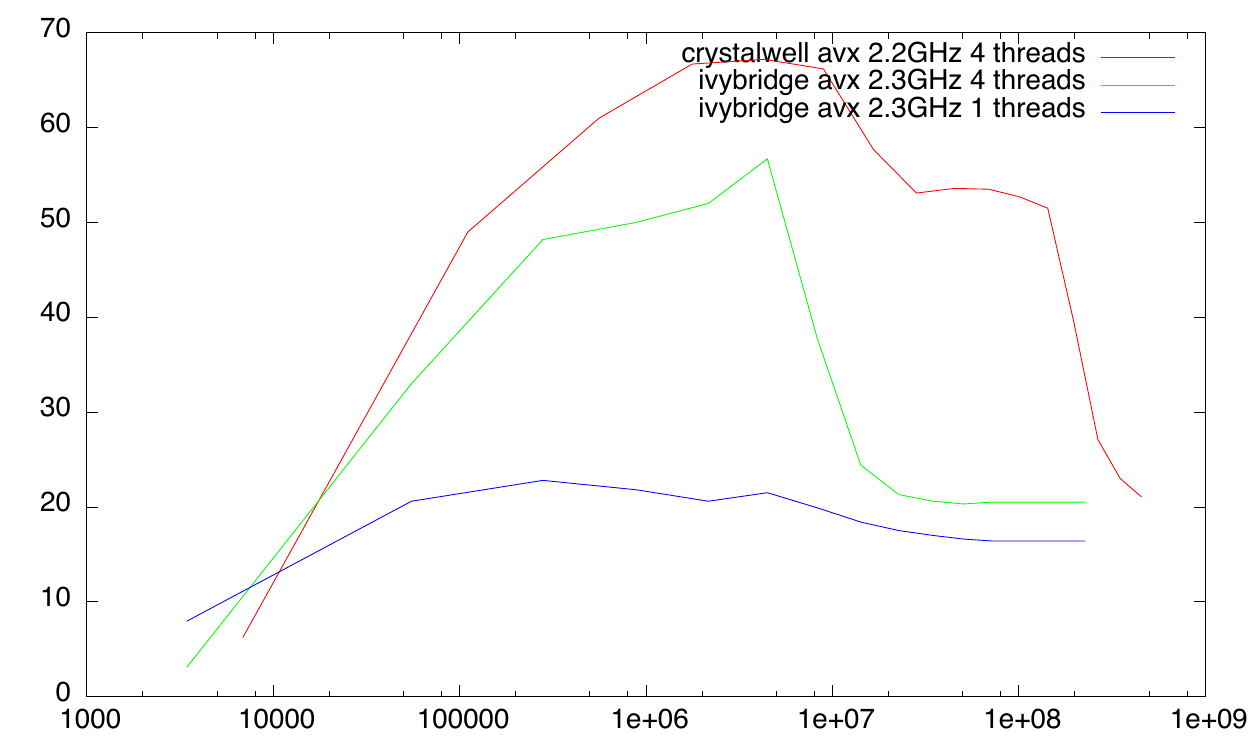}
\caption{\label{fig:cwell} We compare the SU(3)$\times$SU(3) performance (Gflop/s) versus footprint (bytes) 
under AVX-1 instructions of a slightly slower clocked quad-core Haswell (Crystalwell) to a quad-core Ivybridge.
The effect of 128MB integrated on-package eDRAM cache is clearly visible.
}
\end{figure}

Figure~\ref{fig:su3} also displays the performance of Wilson and Domain Wall fermion operators on a quad core Ivybridge. The peak is
$147 GF = 36.8\times 4$ and of this around 40\% was achieved by Grid code. Note that imbalanced multiples and adds
make the maximum achievable 78\%. 

Figure~\ref{fig:cwell} compares the \emph{Crystalwell} variant of Haswell contains a 128MB on package eDram L4 cache to Ivybridge. 
However, to compare the effect of changing only memory performance, we 
used only AVX-1 so that the maximum possible performance of the Haswell system was therefore
actually slightly lower than the Ivybridge. However the benefit of high speed memory integration is then 
clear. 
An AVX-2 compile target is also available
in Grid and does make use of the multiply add instructions where possible. The purpose of this figure is to rather demonstrate the effect
of on-package high bandwidth memory while keeping the available peak performance roughly fixed,
and anticipate the change in performance profile we might expect to see from the Knights Landing chip
which will introduce up to 16GB of very high bandwidth 3D memory on package.

\begin{figure}[hbt]
\includegraphics[width=0.49\textwidth]{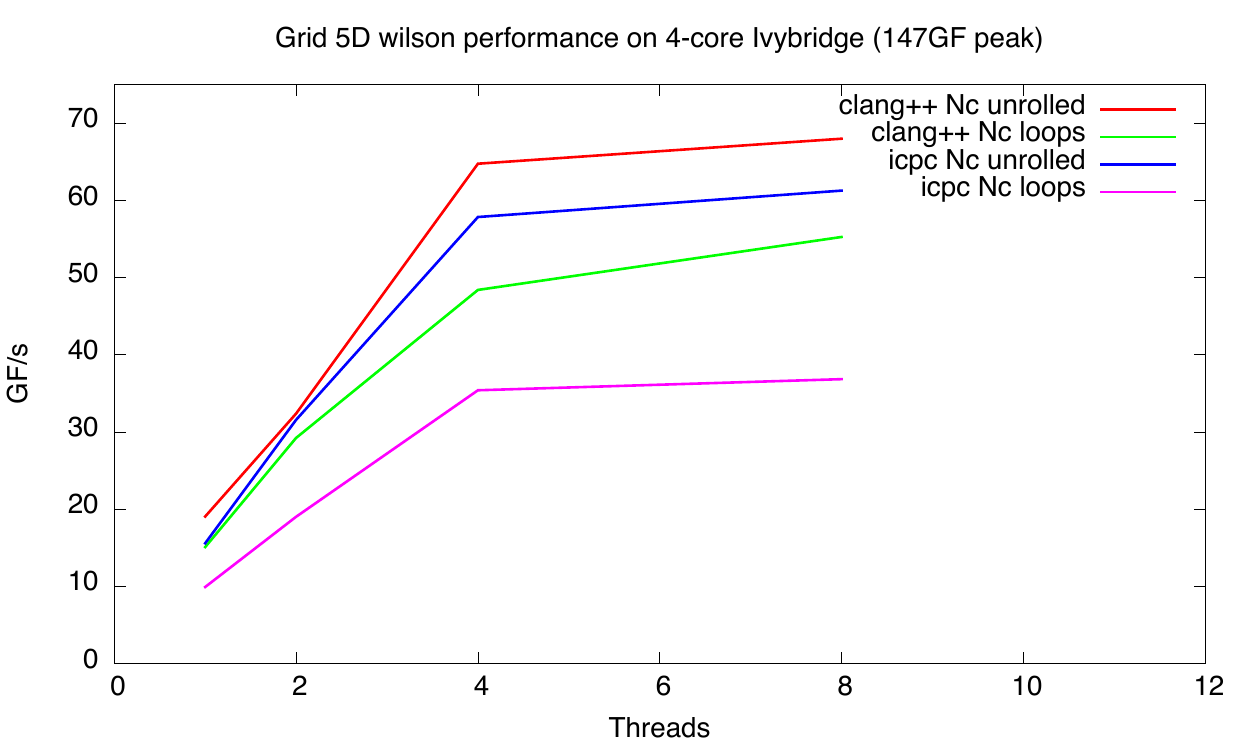}
\includegraphics[width=0.5\textwidth]{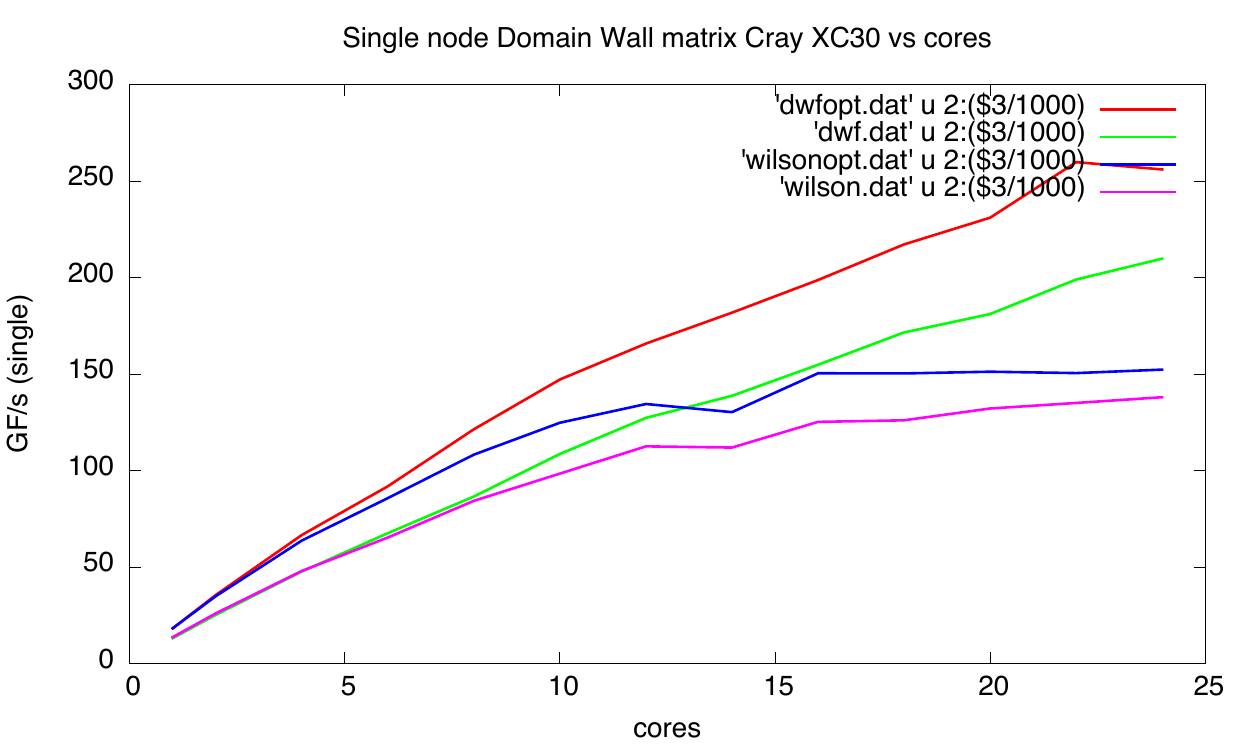}
\caption{\label{fig:xc30} {\bf Left panel:} comparison of the code generation quality for domain wall dslash of
different compilers. A fixed $N_c=3$ version involves hand unrolled using the Grid SIMD data types, while
$N_c$ loops variants work for any $N_c$ but are rather slower for the Intel V16.0 icpc compiler. The Clang++ compiler
performs reasonably even without unrolling.
{\bf Right panel:} The corresponding plot using g++ v4.9 on a 24 core Archer XC30 node. Cache reuse is greater for Domain Wall fermions
than for Wilson Fermions since the gauge field is reused $L_s$ times, while the input vector in the stencil operation is
reused $8$ times. Performance counters suggest the Domain Wall code is very much cache bound and the main memory transfers are roughly
1/8th of the lowest level of cache accesses as expected. }
\end{figure}

Figure~\ref{fig:xc30} demonstrates good performance under
both later versions of GCC, and ICC proving that the performance, although varying a little from compiler to compiler is
portably strong especially when $N_c$ loops are hand unrolled. We compare different generations of Intel and AMD architectures in
Table~\ref{tab:dslash}

\begin{table}[hbt]
\begin{center}
\begin{tabular}{ccc}
Architecture & Cores & GF/s\\
\hline
Cori-Phase1/Haswell   & 32 & 599 \\
Archer/Ivybridge & 24 & 270 \\
BlueWaters/Interlagos & 32 (16) & 80\\
Babbage/Knights Corner & 60 & 270
\end{tabular}
\end{center}
\caption{\label{tab:dslash} Single node benchmark performance of the $L_s$ replicated Wilson dslash routine in single precision on a sample of
current node architectures, applied to an $8^4\times 8$ volume on a single (multi-core) node. We do not factor communication in
because presently the effort is focused on code generation quality. The AMD Interlagos chip has 16 256 bit FPU's shared among 32 Integer
cores making for the core count ``ambiguous''. The Haswell, and Archer systems were dual socket Xeon based, while BlueWaters has a four
socket node. A single socket is used for Knights Corner and we make use of a $16^4 \times 16$ volume since there are so many cores,
and this clearly makes the comparison indirect in this case.}
\end{table}

\section{Status, conclusions and outlook}

Grid is a new and growing physics system and is not yet mature.
The code remains for now a work in progress, and  the development is being performed on a public repository:

\begin{center}
\href{www.github.com/paboyle/Grid}{www.github.com/paboyle/Grid}
\end{center}

Our ultimate focus is to enable the next generation of 5D chiral fermion simulations on the largest systems in the world from 2016-2020 with
outstanding performance. Development is rapid, however and
there is already much of the key functionality in place. Initial performance
figures are very encouraging for a future proof, performance portable, and maintainable code base.

In this paper we have exposed a number of the key programming strategies developed, which we hope reduce
the pain of developing optimised code in future. Significant levels of functionality have already
been implemented and we list these below. The development is rapid, facilitated by the high level programming environment.

\vspace{0.5cm}
\begin{minipage}{0.55\textwidth}
{\bf Algorithms:}
\begin{itemize}
\item Conjugate gradient
\item Conjugate Residual (MCR, GCR)
\item Multi-grid pCG, pGCR
\item HMC, RHMC 
\item Nested integrators \\ (Leapfrog, minimum norm, Force Gradient)
\item Quenched heatbath
\end{itemize}
\end{minipage}
\begin{minipage}{0.44\textwidth}
{\bf Actions:}
\begin{itemize}
\item Wilson Gauge Action
\item Wilson Fermion Action
\item Five dimensional chiral fermions
\item Periodic boundary conditions
\item G-parity boundary conditions
\end{itemize}
\end{minipage}

\section{Acknowledgements}

AY and PB have been supported by Intel through an Intel Parallel Computing Centre held at the Higgs
Centre for Theoretical Physics in Edinburgh. The work has been in part supported by STFC Grants ST/M006530/1,
ST/L000458/1, ST/K005790/1, and ST/K005804/1, by the Grant-in-Aid of the 
Japanese Ministry of Education (No. 15K05065) and by SPIRE (Strategic Program for Innovative Research) Field 5. 
The systems of the STFC funded DiRAC project; Archer at EPCC; Cori, Edison and Babbage at NERSC; and BlueWaters
at NCSA have been used during the code development. System access has been granted by the NERSC
Exascale Science Application Programme (NESAP). We acknowledge useful conversations with the RBC and UKQCD collaborations.

\end{document}